%% file: andris-ssvm21.tex
\documentclass[runningheads]{llncs}
\usepackage{graphicx}
\usepackage{amsmath}
\usepackage{amssymb}
\usepackage{bm}
\usepackage{cite}
\usepackage{tikz}
\usepackage{pgfplots}
\usepackage[mode=buildnew]{standalone}
\usepackage{subfig}

\usetikzlibrary{calc}
\usetikzlibrary{plotmarks}

\begin{document}
	\title{Inpainting-based Video Compression in FullHD
		\thanks{This work has received funding from the European Research Council 
			(ERC) under the European Union's Horizon 2020 research and 
			innovation programme (grant agreement no. 741215, ERC Advanced 
			Grant INCOVID). We thank Matthias Augustin for sharing his in-depth
			knowledge in pseudodifferential inpainting.}
	}
	\author{%Anonymous
		Sarah Andris\inst{1} \and
		Pascal Peter\inst{1} \and
		Rahul Mohideen Kaja Mohideen\inst{1} \and \\
		Joachim Weickert\inst{1} \and
		Sebastian Hoffmann\inst{1}
	}
	\authorrunning{%Anonymous
		S. Andris et al.
	}
	\institute{%Anonymous
		Mathematical Image Analysis Group, \\
		Faculty of Mathematics and Computer Science, \\
		Campus E1.7, Saarland University, 66041 Saarbr\"ucken, Germany \\
		\email{\{andris, peter, rakaja, weickert, hoffmann\}@mia.uni-saarland.de}
	}
	\maketitle              % typeset the header of the contribution
	\begin{abstract}
		Compression methods based on inpainting are an evolving alternative to 
		classical transform-based codecs for still images. Attempts to apply these 
		ideas to video compression are rare, since reaching real-time performance is 
		very challenging. Therefore, current approaches focus on simplified 
		frame-by-frame reconstructions that ignore temporal redundancies.
		As a remedy, we propose a highly efficient, real-time capable prediction and 
		correction approach that fully relies on partial differential equations 
		(PDEs) in all steps of the codec: Dense variational optic flow fields yield 
		accurate motion-compensated predictions, while homogeneous diffusion inpainting 
		is applied for intra prediction. To compress residuals, we introduce a new 
		highly efficient block-based variant of pseudodifferential inpainting. Our 
		novel architecture outperforms other inpainting-based video codecs in terms of 
		both quality and speed. For the first time in inpainting-based video 
		compression, we can decompress FullHD (1080p) videos in real-time with a fully 
		CPU-based implementation, outperforming previous approaches by roughly one 
		order of magnitude. 
		
	\end{abstract}
	\section{Introduction}
	\label{sec:intro}
	
	In today's world, videos are a vital part of our communication, be it personal or professional. Video sharing is constantly rising, making up a large portion of total IP traffic. It is therefore an important task to constantly continue research on improving codecs for video compression. 
	
	Currently, the most well-known and widely used codecs belong to the MPEG family \cite{HPN96}. They are based on hybrid video coding which combines a prediction and a correction step where prediction depends on the frame type. Intra frames rely solely on information from within themselves. Propagating values from preceding or subsequent frames along motion vectors approximates inter frames. The correction step is identical for both types: The residual contains the difference to the original frame and can be efficiently compressed, if the initial prediction was of high quality. Bull \cite{Bu14} gives an approachable introduction into all important ideas leading up to current standards. 
	
	Traditional video codecs use transform-based image compression for 
	residual storage. For still images, however, codecs of this type have 
	been successfully challenged by inpainting-based techniques. They 
	keep values only at a few carefully selected points (the so-called 
	\textit{inpainting mask}) and reconstruct the missing image parts via 
	inpainting. The state of the art for colour images is a PDE-based method by 
	Peter et al.~\cite{PKW16} that extends work by Gali\'c et al.~\cite{GWWB08} and 
	Schmaltz et al.\cite{SPMEWB14} and performs on par with JPEG2000 \cite{TM02} 
	for images with a small to medium amount of texture. As Jost et al.\cite{JPW20} 
	have shown, these methods can even outperform HEVC/intra on piecewise smooth 
	images. 
	
	Classical inpainting problems only reconstruct small amounts of missing image 
	content, but inpainting-based compression relies on sparse known data. 
	Therefore, it tends to be fairly slow, and meeting the real-time requirements of 
	video compression is very challenging. Most existing approaches only focus on 
	specific parts of the video coding pipeline without presenting a full codec 
	\cite{TBS06, LSWZ08, DNLKW10, ZL14, BHR20}. Fully inpainting-based codecs 
	almost exclusively ignore the temporal dimension. They only work on a 
	frame-by-frame basis \cite{KSFR07, PSMMW15} and mainly established speed-up 
	strategies. However, inpainting-based video compression methods will only 
	reach their full potential if they consequently exploit 
	temporal redundancies. So far, only Andris et al.~\cite{APW16} have 
	proposed a modular framework that combines the successful idea of prediction 
	and correction with inpainting-based methods. This framework provides a 
	useful basis for codec design, but the accompanying codec in \cite{APW16} 
	does not use it to its full capacity: It can only reconstruct colour videos 
	with up to a size of $854 \times 364$ in real-time and is limited to small 
	compression ratios due to its simplistic residual storage.
	
	\subsection{Our Contribution}
	\label{sec:contribution}
	
	Our goal is to design a codec with significantly improved 
	performance both in quality and speed compared to \cite{APW16} and other 
	inpainting-based codecs. To this end, we use the framework from \cite{APW16}
	as a basis and consequently implement 
	inpainting-based methods for all prediction and compression submodules. Our 
	hybrid inpainting-based video codec (HIVC) employs the methods of Brox et 
	al.~\cite{BBPW04} for optic flow field computation, homogeneous diffusion 
	inpainting \cite{Ca88} for intra prediction, and finite state entropy (FSE) 
	coding \cite{FSE}. For representing the residuals, we design a block-based 
	variant of pseudodifferential inpainting \cite{AWA19} which we can fully 
	describe in fast cosine transforms on the decoder side. Through efficient 
	algorithms, we are able to push real-time reconstruction of colour videos to 
	FullHD 1080p resolution without resorting to parallelisation on the GPU. 
	Furthermore, we can reach compression ratios which are over fifty times larger 
	than in \cite{APW16} while even slightly increasing quality.
	
	\subsection{Related Work}
	\label{sec:related}
	
	Inpainting-based techniques have been applied for several parts of the coding 
	pipeline or even for complete codecs. K\"ostler et al.~\cite{KSFR07} were the 
	first to achieve real-time performance with a codec based on homogeneous 
	diffusion inpainting on a \textit{Playstation 3}. The state of the art in 
	PDE-based image compression, the R-EED codec by Schmaltz et 
	al.~\cite{SPMEWB14}, has been adapted by  Peter et al.~\cite{PSMMW15} for video 
	compression. However, these codecs work on a strict frame-by-frame basis. First 
	attempts to incorporate motion information were made by Schmaltz and Weickert 
	\cite{SW12} who combine pose tracking with static background compression by 
	anisotropic diffusion inpainting. Breu{\ss} et al.~\cite{BHR20} still perform 
	frame-based inpainting, but they acquire masks by shifting one optimised mask 
	along motion vectors. They do not present a full video compression pipeline. 
	Another stand-alone codec is proposed by Wu et al.~\cite{WSK18} who employ deep 
	learning to interpolate inter frames between two intra frames. %Their network 
	%takes motion fields into account and codes residuals, which makes their method 
	%closely related to hybrid video coding. 
	Liu et al.~\cite{LSWZ08}, Doshkov et al.~\cite{DNLKW10}, and Zhang and Lin \cite{ZL14} incorporate inpainting ideas directly into the intra prediction of a hybrid video coder. Their methods combine homogeneous diffusion inpainting with edge information, template matching, and adaptive boundary values, respectively. For the same task, Tan et al.~\cite{TBS06} perform inpainting via template matching. Jost et al.~\cite{JPW20} focus on the compression of dense optic flow fields and design a well-performing method employing edge-aware homogeneous diffusion inpainting. However, their method is not real-time capable.
	%In the field of video completion, which tries to fill in missing regions e.g. for object removal, most papers are only connected to our work via the notion of inpainting. However, El Helou et al.~\cite{EZSG+20} drive their method to the extreme by using a very sparse set of known data (1-2\%) and applying the well-known Shepard interpolation \cite{Sh68}, \cite{AAS17} to reconstruct regions in between.  
	
	Since high quality motion prediction is essential for efficient video 
	compression, dense optic flow fields have been investigated in multiple works. 
	Li et al. \cite{LHX18} employ the classical Horn and Schunck \cite{HS81} 
	approach for bidirectional prediction. More complex techniques such as Bayesian 
	methods in Han and Podilchuk \cite{HP01} and velocity field modeling in Chen 
	and Mied \cite{CM13} yield optic flow fields with higher prediction quality. 
	Both methods take also the compressibility of the acquired motion into account. 
	Going further into this direction, Ottaviano and Kohli \cite{OK13} represent 
	optic flow in a wavelet basis and obtain the corresponding coefficients by 
	minimising the residual after inter prediction. All of these works consider 
	flow fields independent of compression or as an augmentation for 
	transform-based video codecs. In contrast, we embed them into a fully 
	inpainting-based approach that addresses all relevant coding steps.

	\subsection{Paper Structure}
	\label{sec:structure}
	
	We present our video codec in Section \ref{sec:codec} and discuss corresponding experiments in Section \ref{sec:experiments}. We conclude our paper in Section \ref{sec:conclusion}.
	
	% ------------------------------------------------------------------------------
	% Codec
	
	\section{A Fully Inpainting-Based Video Codec}
	\label{sec:codec}
	
	We generalise the framework by Andris et al.~\cite{APW16}, but keep the same 
	central concept of prediction and correction. In a first step, we compute and 
	subsequently compress dense \textit{backwards optic flow fields (BOFFs)} 
	between frames. We represent one scene by a \textit{group of pictures (GOP)}, 
	which includes one intra frame at the beginning and subsequent inter frames. The size of these GOPs may vary according to video content. In 
	a second step, we predict intra frames with some inpainting technique and inter 
	frames with motion compensation based on the previously computed BOFFs. For 
	each frame, a corresponding correction (residual) contains the difference to 
	the originals. These residuals, again, have to be compressed. Finally, we store 
	and encode all data needed for decoding.
	
	In the following, we present a concrete codec implementation for this general framework. The most important ingredients are the intra prediction and residual compression techniques, which we investigate in Section \ref{sec:hominp} and \ref{sec:pseudo}, respectively. We then show our final codec design with a description of all relevant components in Section \ref{sec:modeling}.
	
	\subsection{Global Homogeneous Diffusion Inpainting}
	\label{sec:hominp}
	
	Let $\bm f$ be a 1D representation of a 2D discrete image of size $n_x \times n_y$, i.e. we sort the image pixels row-wise into the vector $\bm f \in \mathbb{R}^N$ with $N=n_xn_y$. We store values only at a few selected locations represented by the binary inpainting mask $\bm m \in \mathbb{R}^N $ and discard all other values. Then, we can acquire an approximation $\bm u \in \mathbb{R}^N$ of the original image by solving the general discrete inpainting problem
	\begin{align}
	\bm{M} (\bm u - \bm f) - (\bm I - \bm M) \bm A \bm u = \bm 0 \, .
	\end{align}
	The matrix $\bm M\in \mathbb{R}^{N\times N}$ contains the entries of the inpainting mask $\bm m$ on its diagonal, and is zero everywhere else. $\bm I$ is the identity matrix, and $\bm A$ represents a discrete inpainting operator with reflecting boundary conditions. The first term ensures that $\bm u$ adopts the original values from $\bm f$ at mask positions, the second term realises inpainting steered by the operator $\bm A$ in regions inbetween. 
	%Designing $\bm A$ is a crucial task in codec design. 
	The choice of $\bm A$ influences the reconstruction quality immensely and simultaneously affects the complexity of the algorithms solving the inpainting problem.
	%We opt for $\bm A = - \bm L$ with $\bm L$ being a discrete Laplacian. 
	Choosing $\bm A = - \bm L$ with discrete Laplacian $\bm L$ results in 
	homogeneous diffusion inpainting \cite{Ca88}, which presents a good balance 
	between quality and simplicity.
	
	We use a coarse-to-fine algorithm similar to the cascadic conjugate gradient 
	method by Deuflhard \cite{De94} to solve the arising system of equations. The 
	basic idea is to build an image pyramid by subsampling, solving the system on a 
	coarse level, and use this solution as input for the next finer level. This 
	yields a much faster convergence (both on the individual levels and in total) 
	compared to classically solving the system only on the finest level.

	\subsection{Block-based Pseudodifferential Inpainting}
	\label{sec:pseudo}
	
	Pseudodifferential inpainting has been established by Augustin et al.~\cite{AWA19} as a connecting concept between inpainting with rotationally invariant PDEs and radial basis function (RBF) interpolation. Their paper extends results of Hoffmann et al.~\cite{HPW15} on harmonic and biharmonic inpainting with Green's functions. We use their work to build a highly efficient algorithm for inpainting based on the discrete cosine transform, which yields major improvements for the final codec compared to \cite{APW16}. For images, mask, and operator, we stick to the notation used in Section \ref{sec:hominp}.
	
	Introducing the theoretical background of Green's functions is beyond the scope of this paper. We refer the interested reader to \cite{HPW15}. Instead, we provide an intuitive interpretation: For a discrete symmetric inpainting operator $\bm A$, the corresponding discrete Green's function $\bm g_{k}$ characterises the influence of an impulse at pixel position $k$. Instead of solving a system of equations for the general inpainting problem, we can then directly obtain a solution via a weighted sum of the operator's Green's functions at mask positions: 
	%the inner product of a coefficient vector $\bm c$ with the operator's Green's functions at mask positions. We 
	\begin{equation}
	\label{eq:greens1}
	%u_k = \langle \bm c, \bm g_k \rangle + a
	u_k = \sum_{i=1}^{n_xn_y} \left( m_i \cdot c_i \cdot (\bm g_i)_k \right) + a \, \, \, \, \, \text{ for } k = 1, ..., N.
	%u_{i,j} = \sum_{(k,l) \in K} c_{k,l} \cdot (\bm g_{k,l})_{i,j} + a \, .
	\end{equation}
	Recall that $\bm m=(m_i)_{i=1,...,N}$ is the inpainting mask and $\bm 
	u=(u_i)_{i=1,...,N}$ the corresponding inpainting solution in vector notation. 
	The coefficient vector $\bm c$ and the constant $a$ can be acquired by solving 
	a linear system of equations of size $(K+1) \times (K+1)$, where $K$ is the 
	number of mask points; see \cite{HPW15} for more details. In contrast to the 
	sparse but large matrix used for solving the general inpainting problem 
	directly, the system matrix is fairly small and densely populated.
	
	It is well-known that the discrete Green's functions build up the pseudo-inverse of the corresponding operator. Therefore, they are symmetric, i.e. $(\bm g_i)_k = (\bm g_k)_i$ for all $i, k$, since we assumed $\bm A$ to be symmetric. We define $\bm G$ as the matrix containing the Green's functions $\bm g_k$ as its columns. Then we can reformulate Equation (\ref{eq:greens1}) as
	\begin{equation}
	\label{eq:greens2}
	\bm u = \bm G \bm M \bm c + \bm a
	\end{equation}
	where the vector $\bm a$ has the constant $a$ in every entry. Since our 
	inpainting operator $\bm A$ is a finite difference matrix and the Green's 
	functions are its pseudo-inverse, $\bm G$ is also a difference matrix. Hence, 
	we can apply results by Strang and MacNamara \cite{SM14}, proving that $\bm G$ 
	is Toeplitz-plus-Hankel. Sanchez et al. showed in \cite{SGPS+95} that matrices 
	of this type are diagonalisable by the even discrete cosine transform (DCT) of 
	type II. Denoting the corresponding transform matrix by $\bm{\mathcal{C}}$, we 
	can rewrite Equation (\ref{eq:greens2}) as
	\begin{align}
	\bm u & = \bm{\mathcal{C}}^{-1} \, \text{diag}(\bm \lambda) \, \bm{\mathcal{C}} (\bm M \bm c) + \bm a  
	\end{align}
	where the vector $\bm \lambda$ contains the eigenvalues of the Green's functions. Thus, if we have the coefficients $\bm c$, we can acquire the final inpainting solution by multiplying the coefficients with the eigenvalues in the transform domain, computing the backtransform, and adding the constant to all pixels.
	
	All these considerations hold for arbitrary image sizes. If we now subdivide the image domain into $8 \times 8$ blocks and perform inpainting on each block independently, we can employ a dedicated fast DCT algorithm. We opt for the method by Arai et al.~\cite{AAN88} which is also used in JPEG. We treat each block as an independent image, i.e.~impose mirrored boundary conditions. Note that if we store the coefficients and the constant on the encoder side, the decoder only has to perform two fast DCTs, $n_xn_y$ multiplications, and $n_xn_y$ additions to acquire the final inpainting result. This brings our method close to the main concept of JPEG. However, in contrast to JPEG, our coefficients do not correspond to frequencies, but give a connection to local structures. Moreover, the inpainting operator is now completely defined by the eigenvalues of the Green's functions and can be easily replaced without changing the algorithm or influencing the speed on the decoder side.
	
	\begin{figure*}[!t]
		\centerline{
			\subfloat[Original (intra) frame 1853]{
				\includegraphics[width=0.4\textwidth]{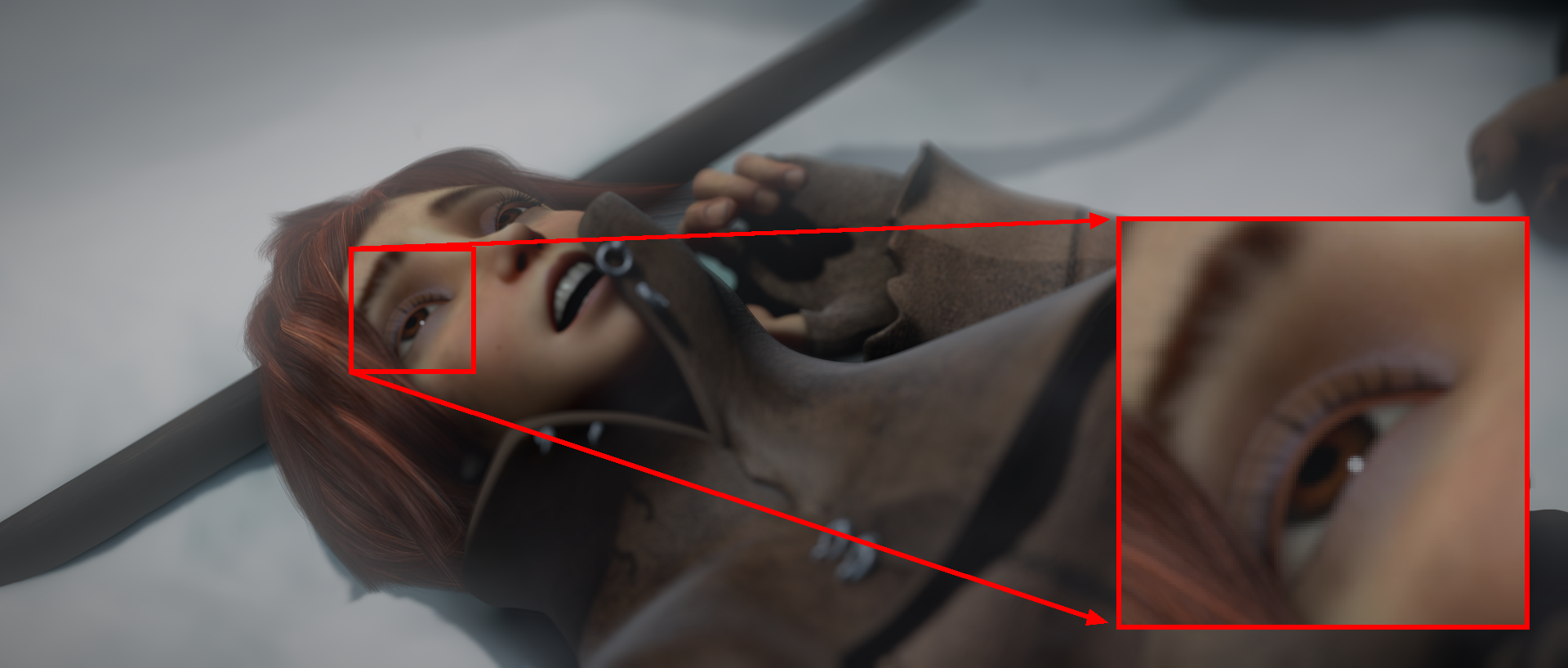}%
				\label{fig:intra-orig}
			}
			\hfil
			\subfloat[Inpainting mask Y-channel.]{
				\includegraphics[width=0.4\textwidth]{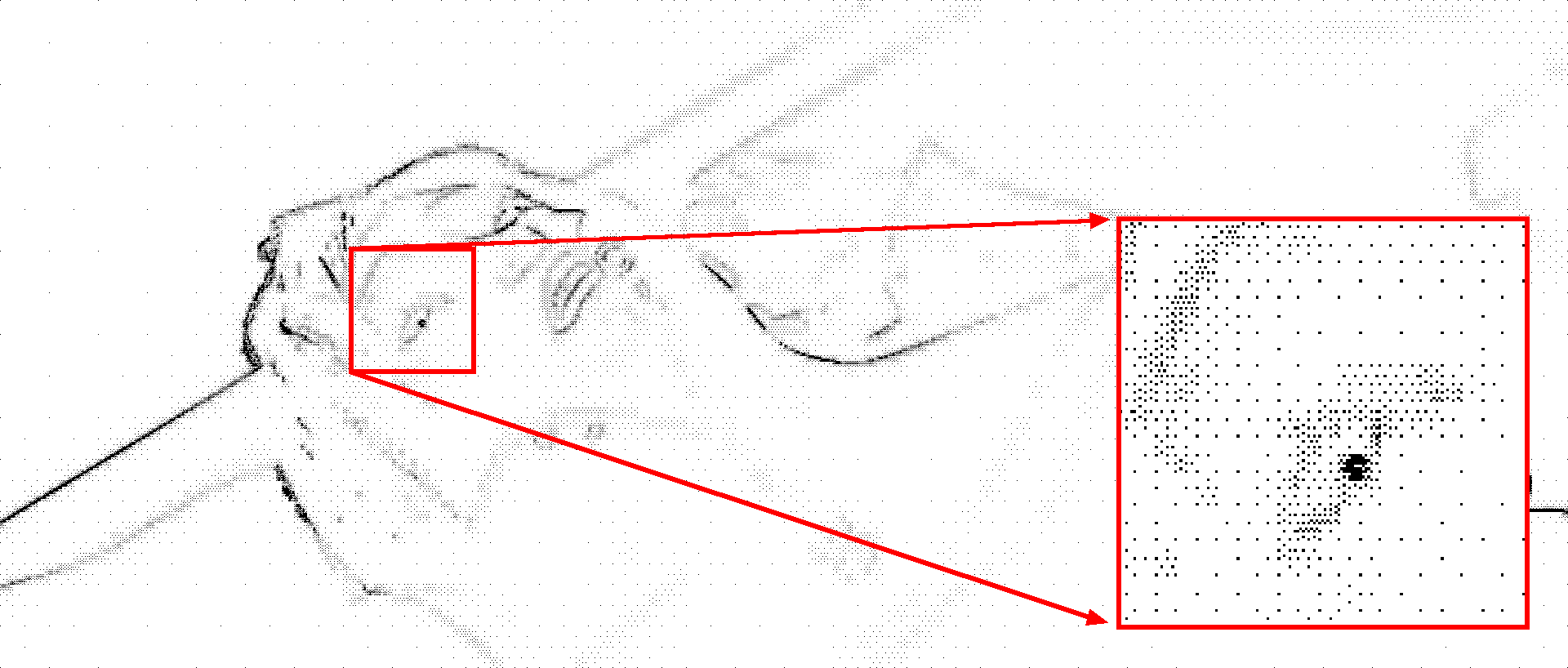}%
				\label{fig:intra-mask-Y}
			}
		}
		\centerline{
			\subfloat[Inpainting mask UV-channels.]{
				\includegraphics[width=0.4\textwidth]{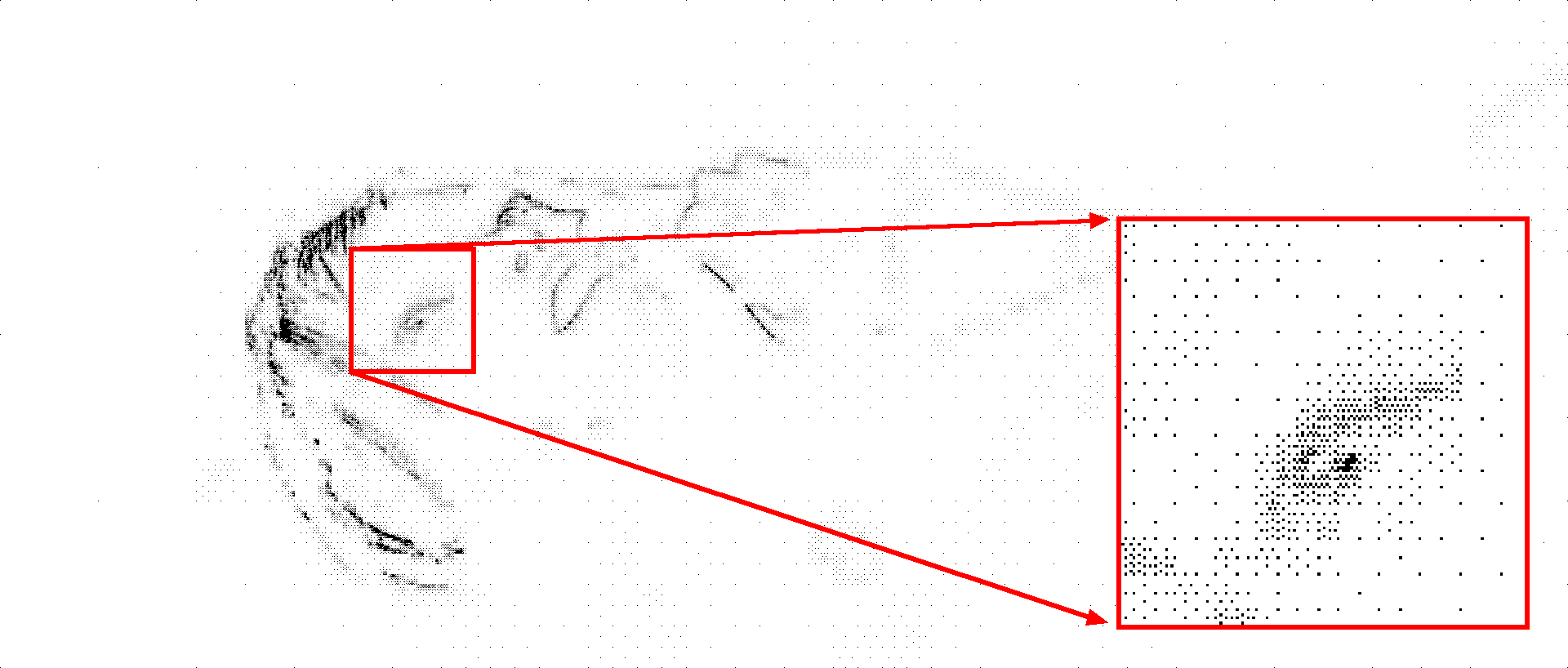}%
				\label{fig:intra-mask-CbCr}
			}
			\hfil
			\subfloat[Prediction.]{
				\includegraphics[width=0.4\textwidth]{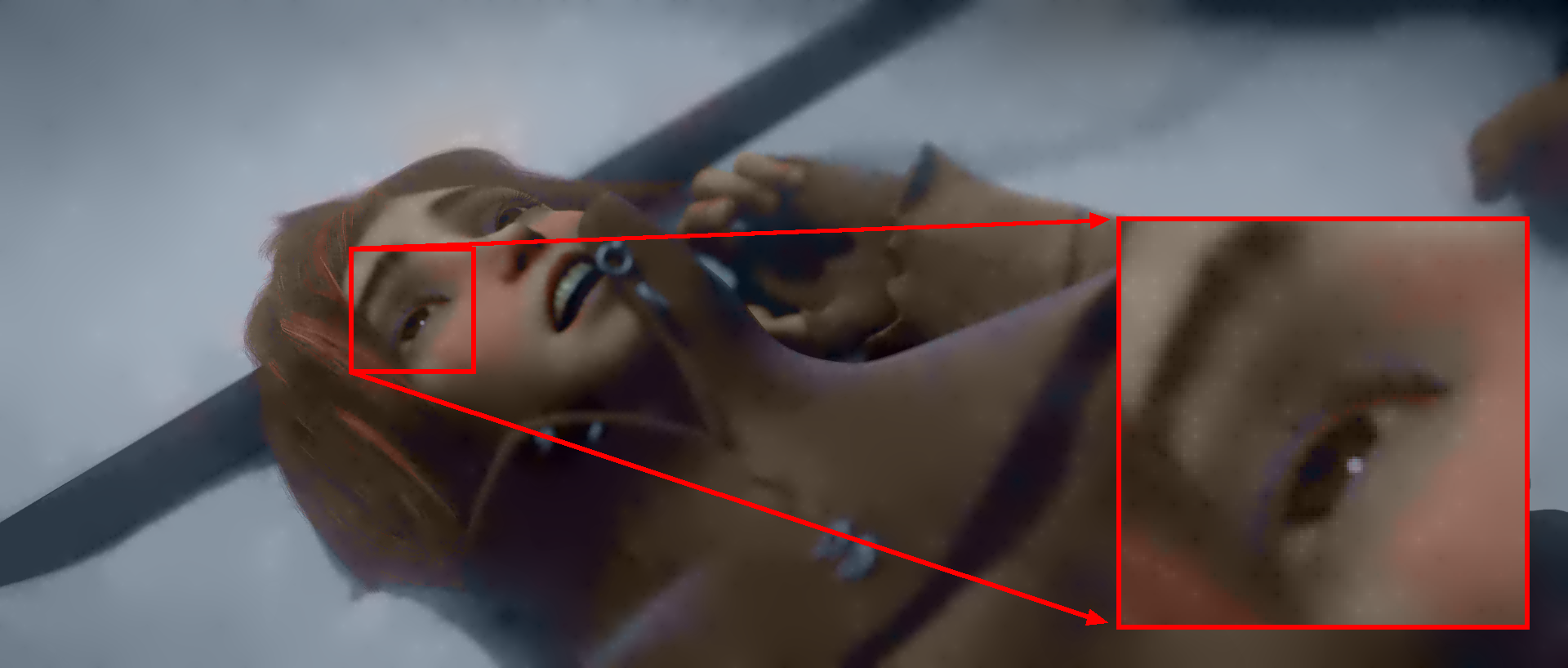}%
				\label{fig:intra-predic}
			}
		}
		\centerline{
			\subfloat[Residual.]{
				\includegraphics[width=0.4\textwidth]{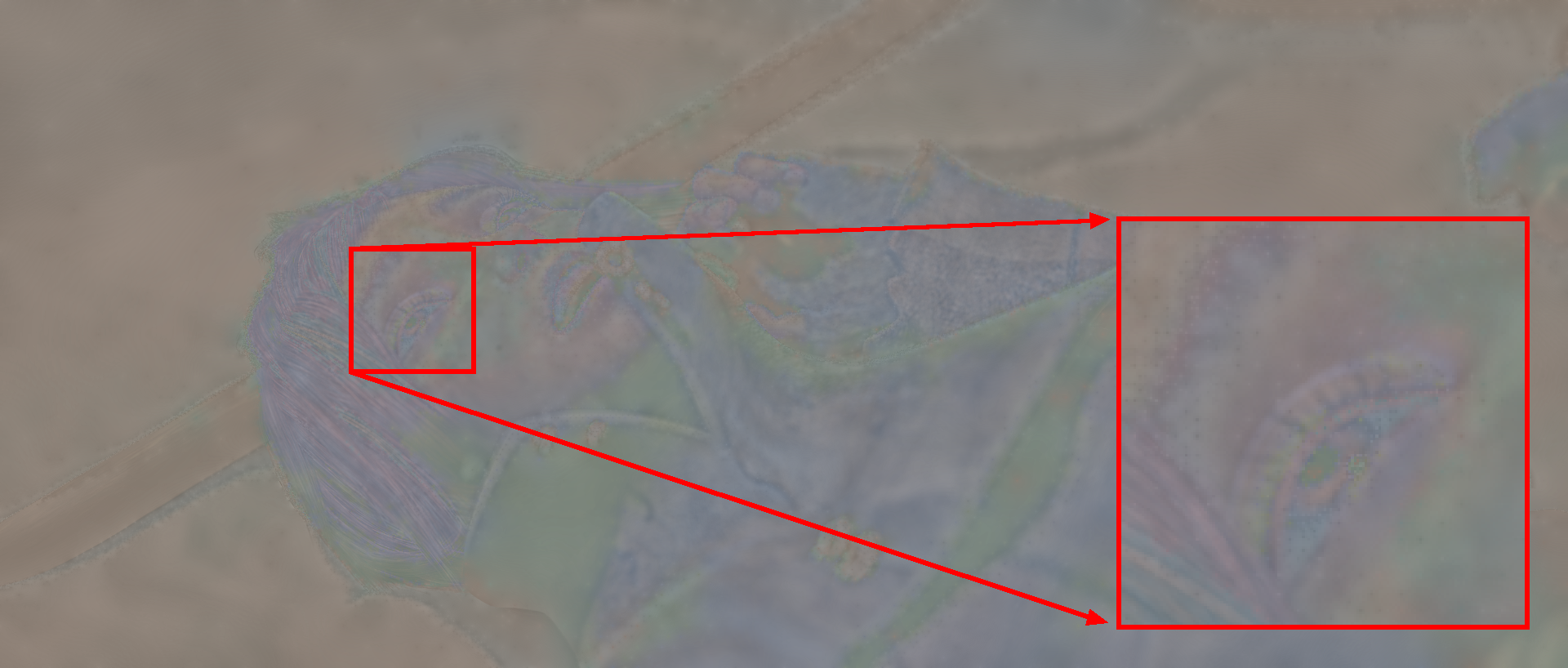}%
				\label{fig:intra-res}
			}
			\hfil
			\subfloat[Final Reconstruction.]{
				\includegraphics[width=0.4\textwidth]{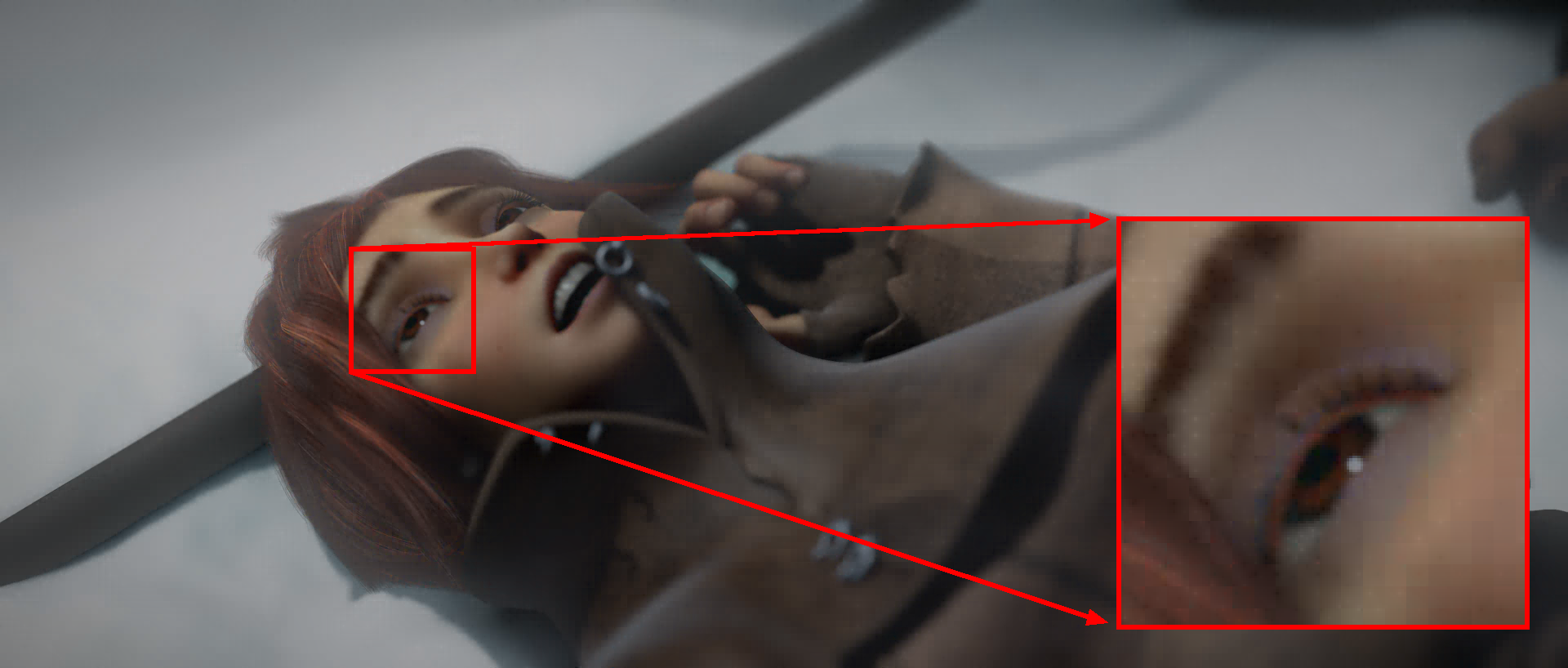}%
				\label{fig:intra-recon}
			}
		}
		\caption{Intermediate results for intra coding at a compression rate of roughly 100:1. Global homogeneous diffusion inpainting is especially suited for smooth regions. Thus, the mask concentrates at edges. With the residual we can correct remaining errors.}
		\label{fig:intra}
	\end{figure*}
	
	\begin{figure*}[!t]
		\centerline{
			\subfloat[Original (inter) frame 1854.]{
				\includegraphics[width=0.4\textwidth]{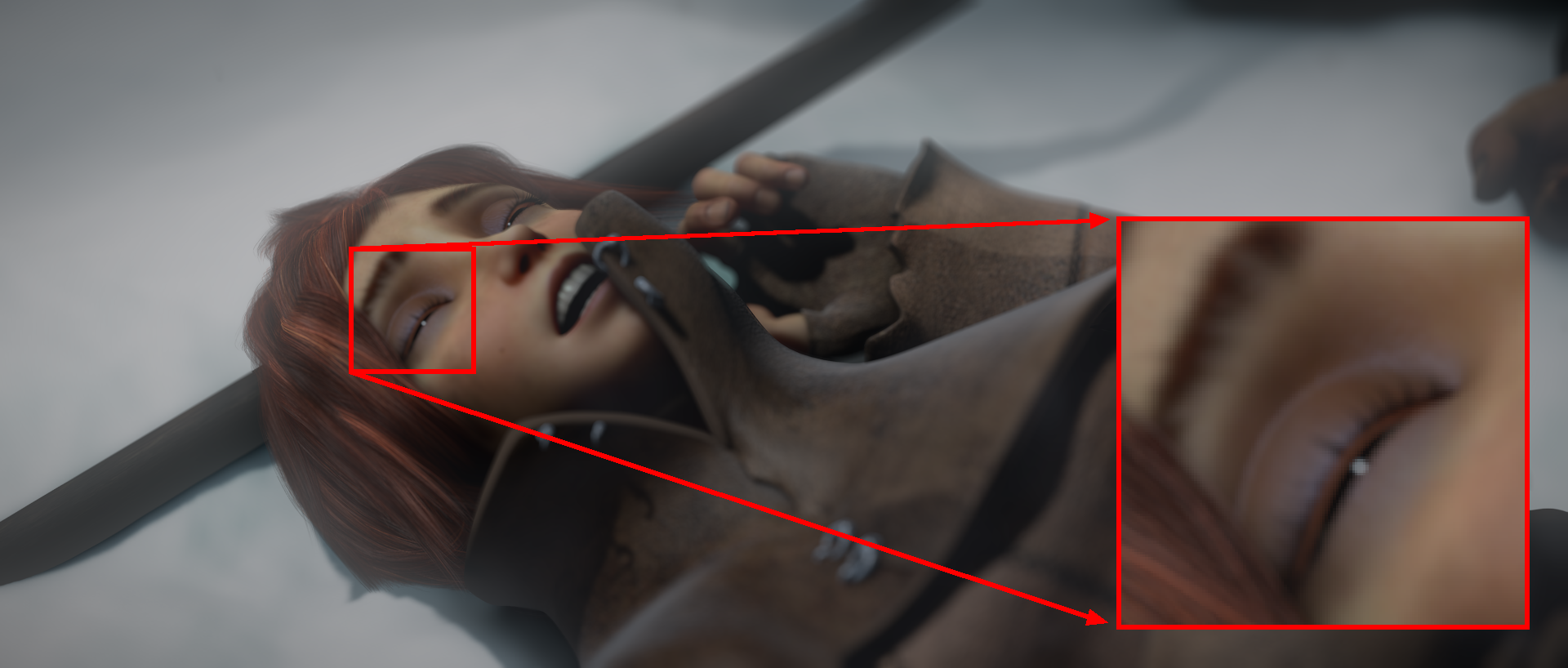}%
				\label{fig:inter-orig}
			}
			\hfil
			\subfloat[BOFF.]{
				\includegraphics[width=0.4\textwidth]{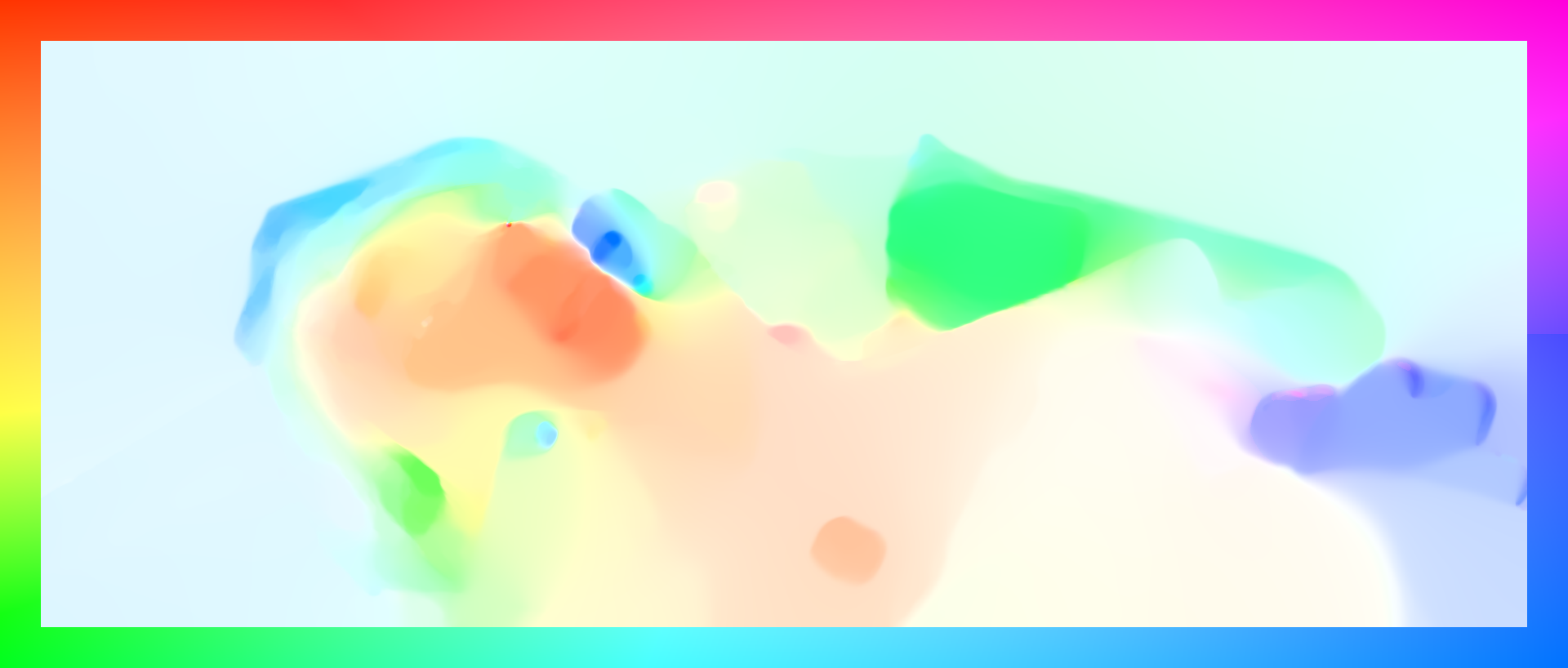}%
				\label{fig:inter-flow}
			}
		}
		\centerline{
			\subfloat[Compressed BOFF.]{
				\includegraphics[width=0.4\textwidth]{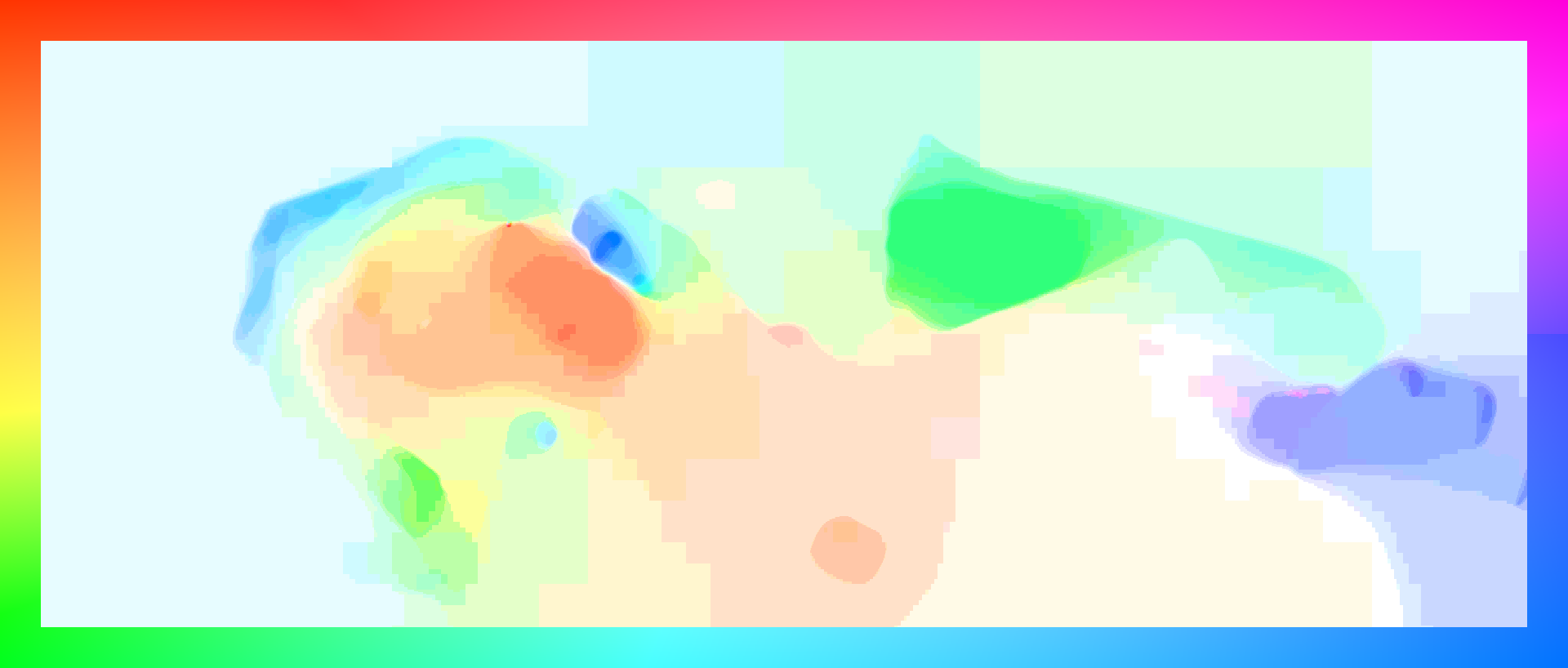}%
				\label{fig:inter-flowcomp}
			}
			\hfil
			\subfloat[Prediction.]{
				\includegraphics[width=0.4\textwidth]{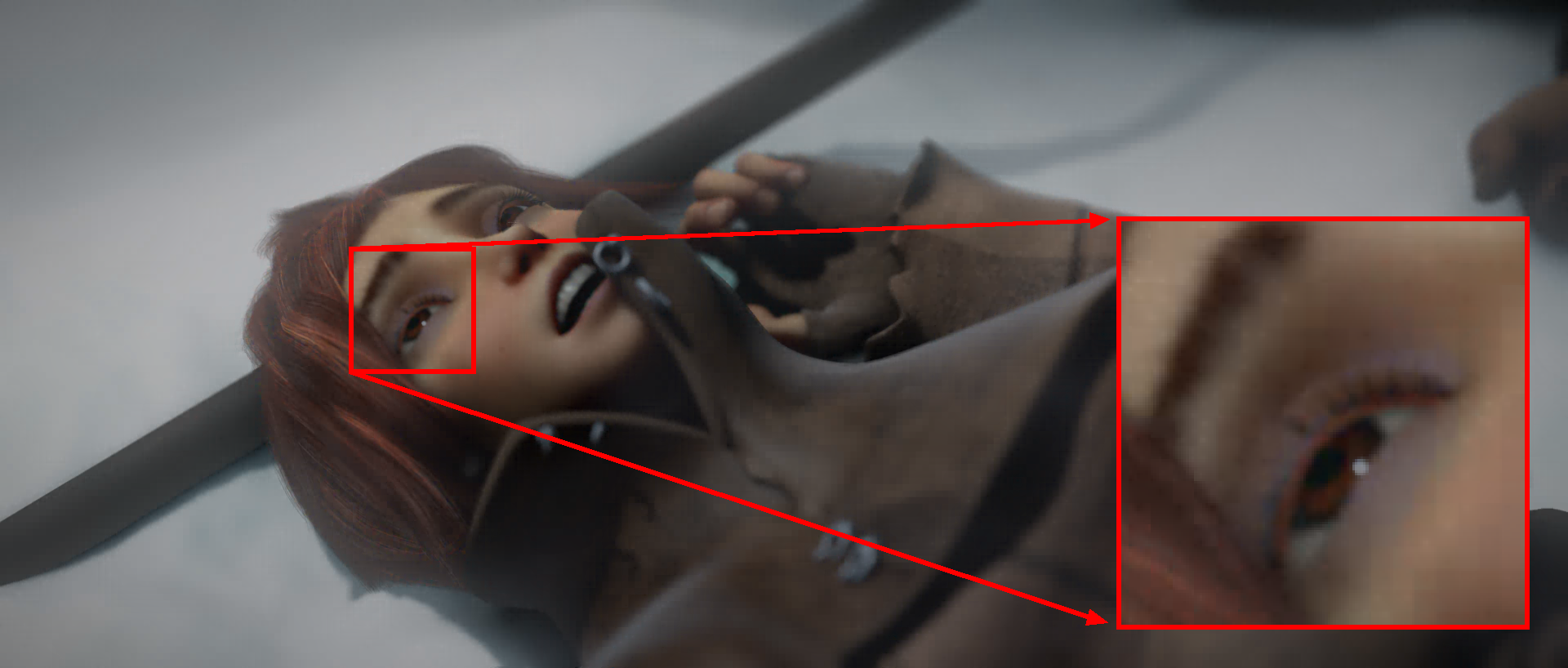}%
				\label{fig:inter-predic}
			}
		}
		\centerline{
			\subfloat[Residual.]{
				\includegraphics[width=0.4\textwidth]{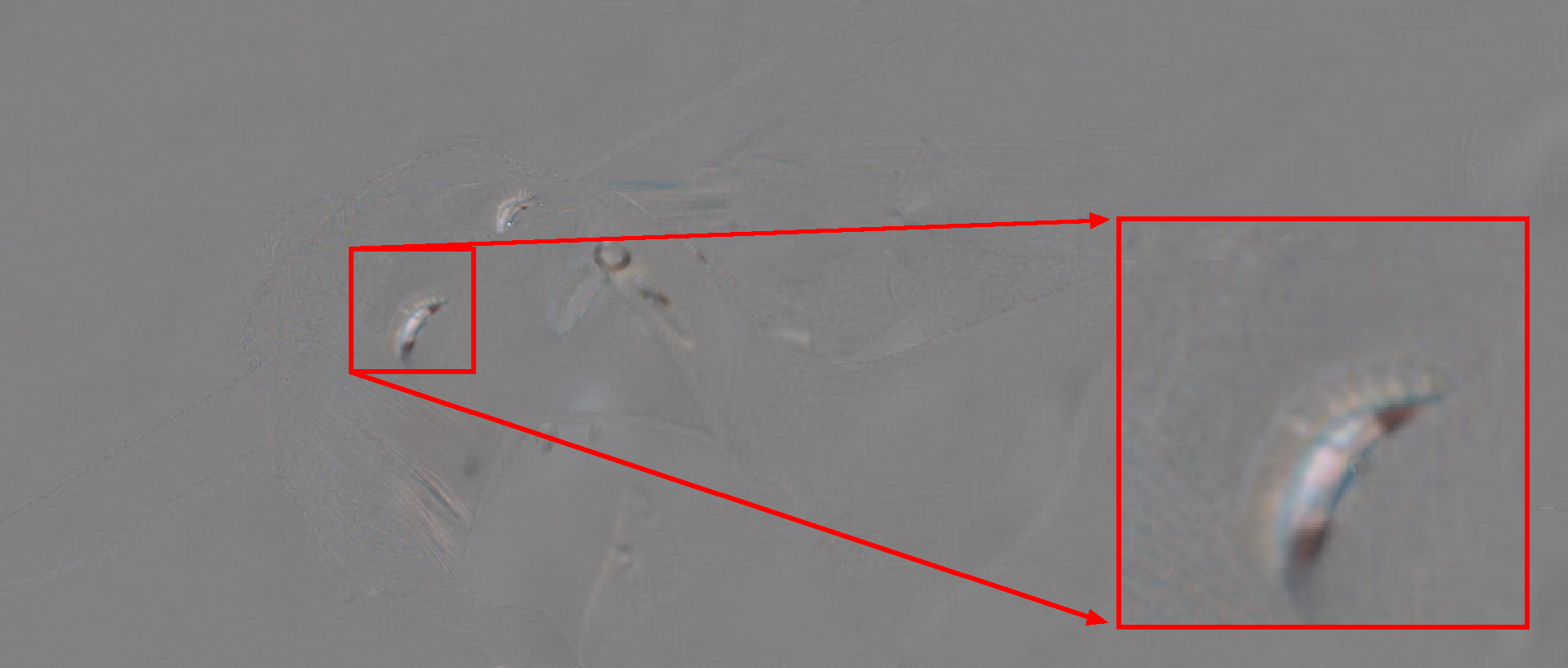}%
				\label{fig:inter-residual}
			}
			\hfil
			\subfloat[Final Reconstruction.]{
				\includegraphics[width=0.4\textwidth]{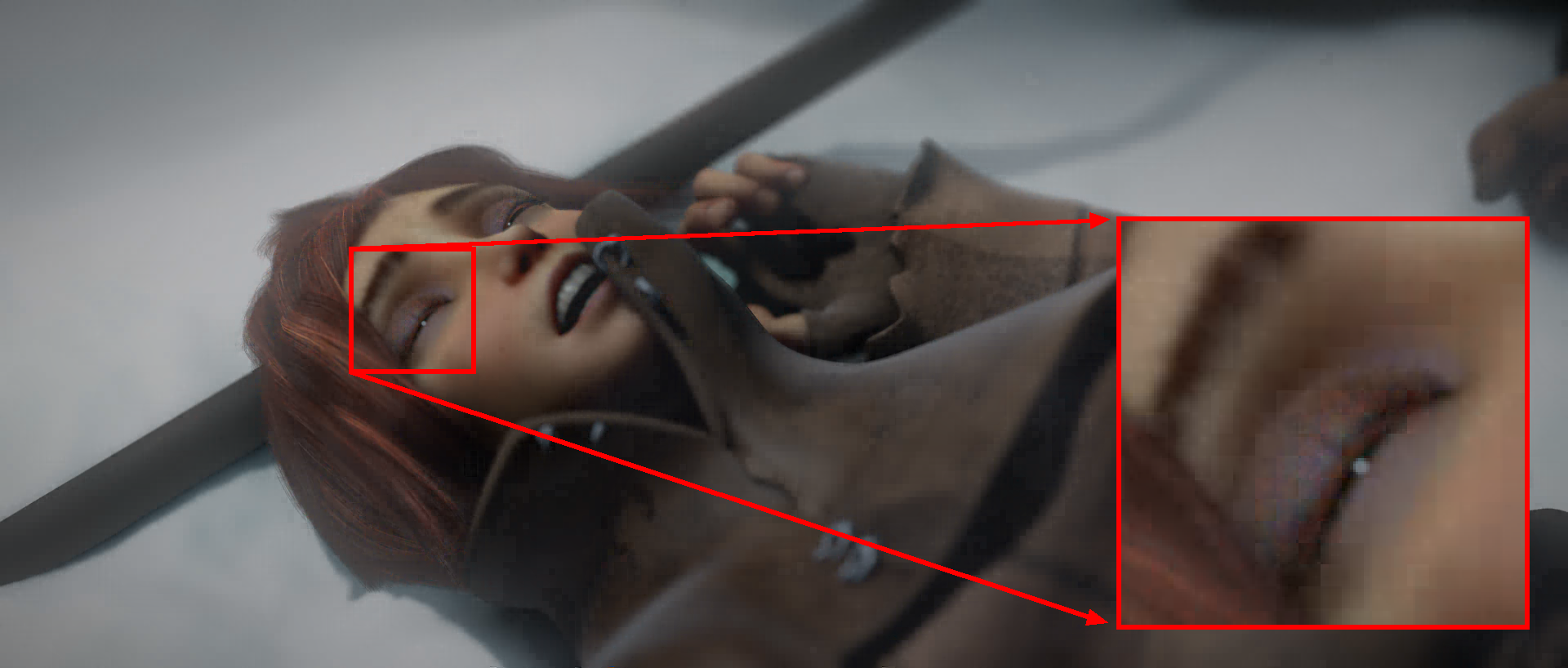}%
				\label{fig:inter-recon}
			}
		}
		\caption{Intermediate results for inter coding at a compression rate of roughly 100:1. The colour coding of the BOFF is adapted from \cite{BSLRBS11}. The flow field yields a prediction of high accuracy and only struggles at occlusions and disocclusions (e.g.~the closing eyes).}
		\label{fig:inter}
	\end{figure*}
	
	\begin{figure*}[!t]
		\centerline{
			\subfloat[Residual for frame 1853.]{
				\includegraphics[width=0.4\textwidth]{Pics/visual/intra-res}%
				\label{fig:res-orig}
			}
			\hfil
			\subfloat[Inpainting mask Y-channel.]{
				\includegraphics[width=0.4\textwidth]{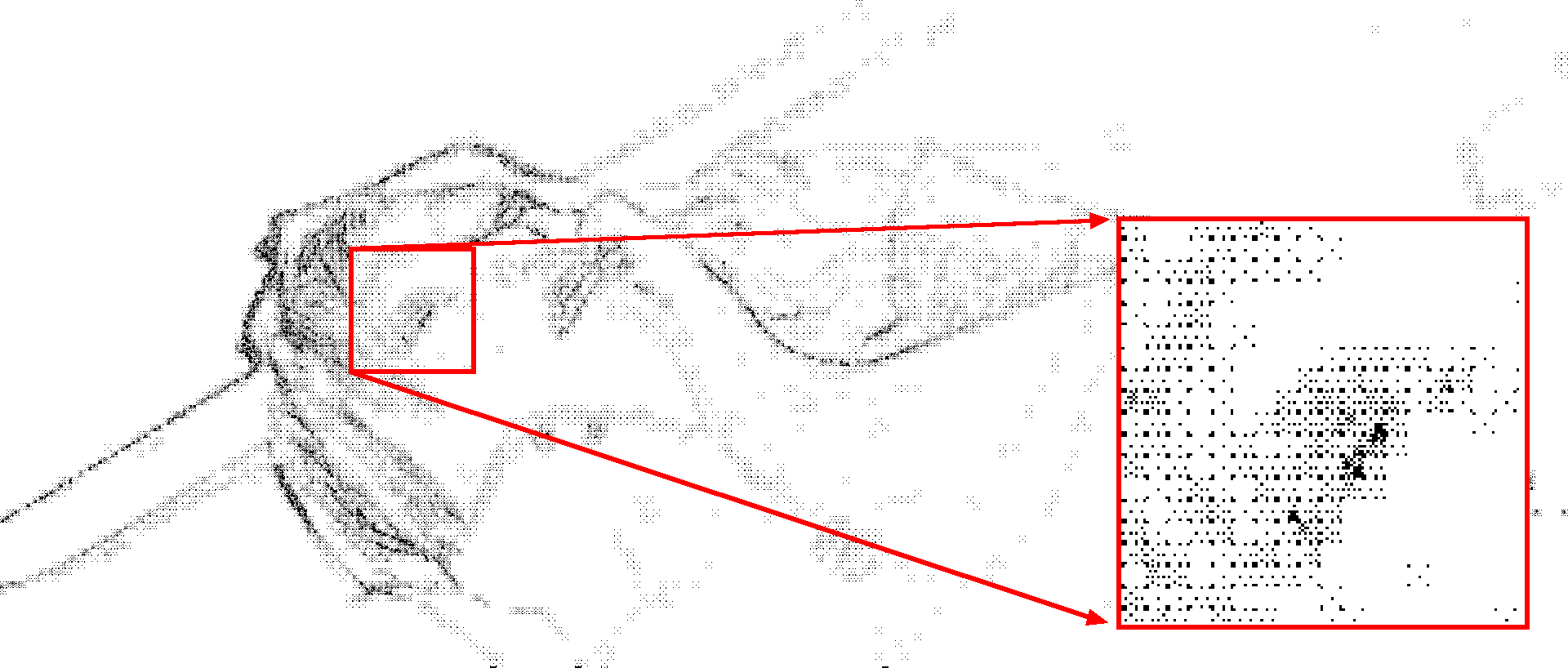}%
				\label{fig:res-mask-y}
			}
		}
		\centerline{
			\subfloat[Inpainting mask UV-channels.]{
				\includegraphics[width=0.4\textwidth]{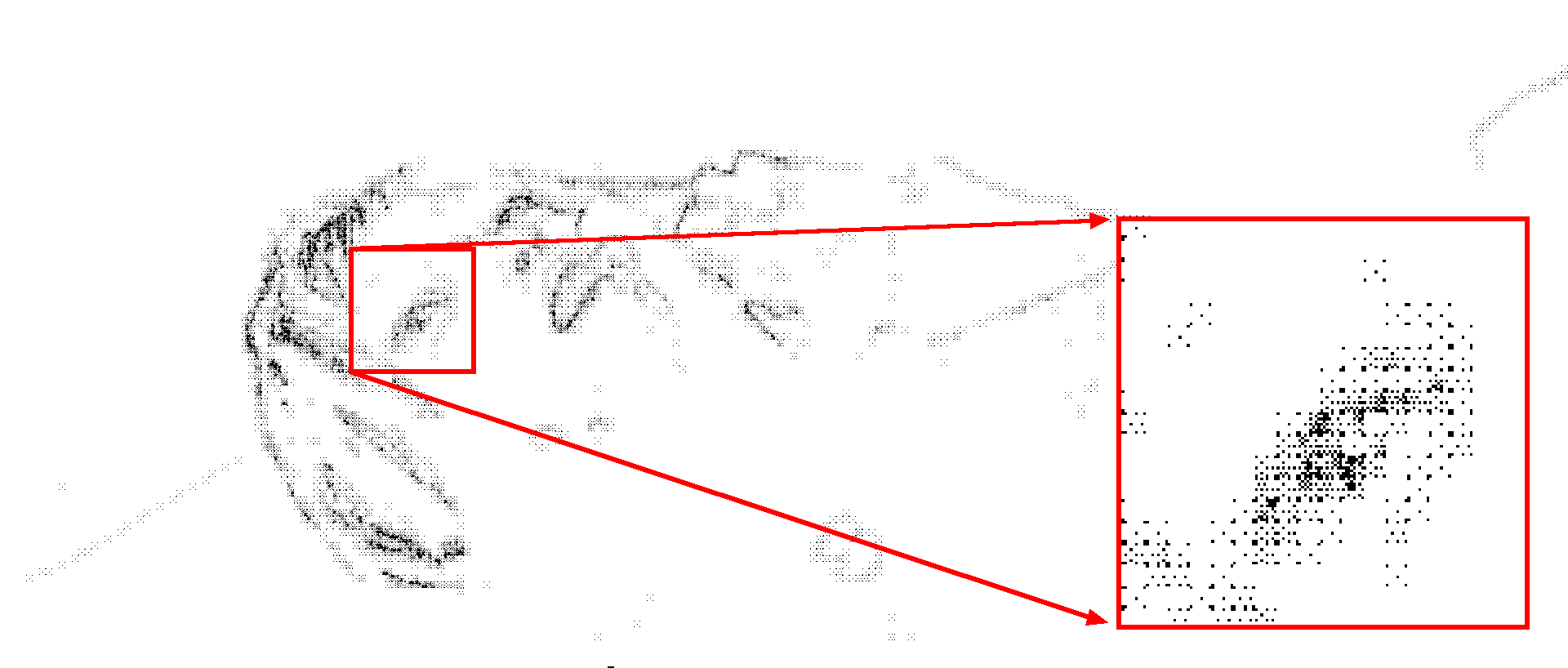}%
				\label{fig:res-mask-cbcr}
			}
			\hfil
			\subfloat[Compressed Residual.]{
				\includegraphics[width=0.4\textwidth]{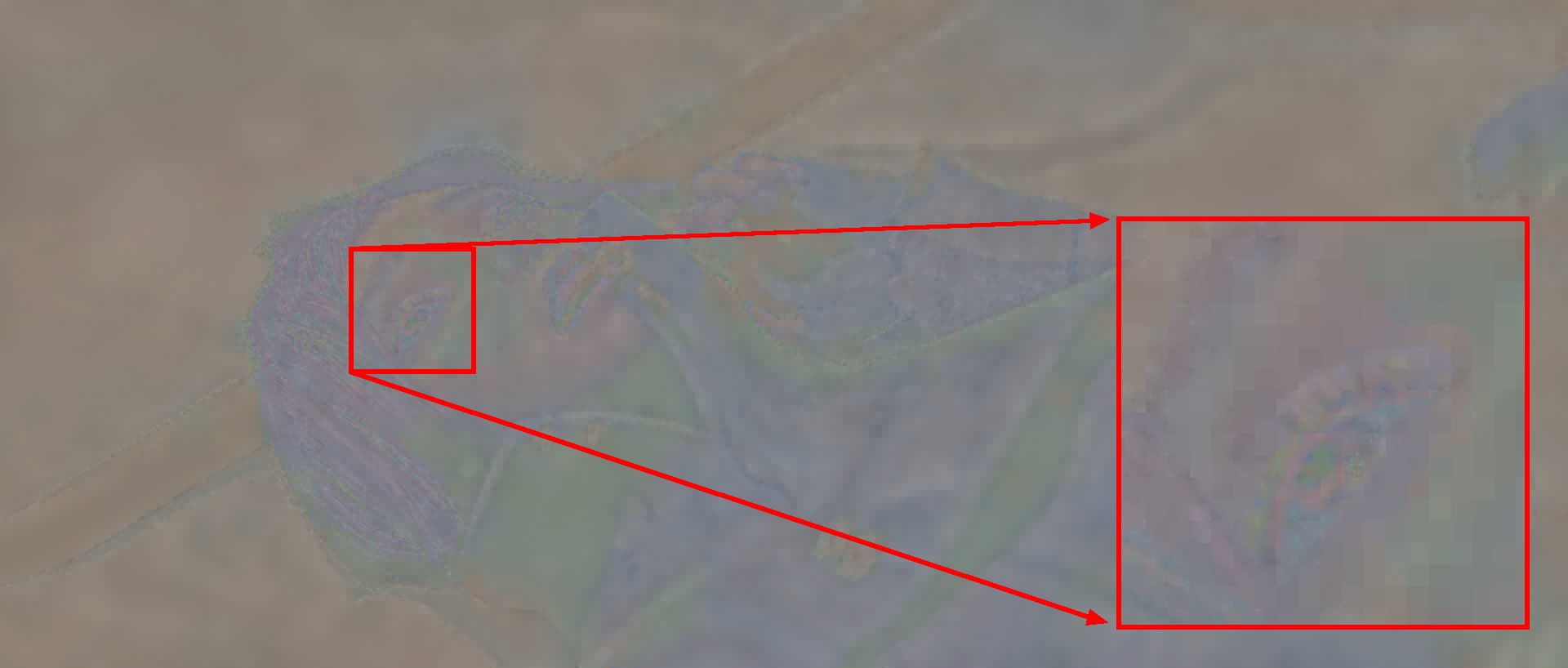}%
				\label{fig:res-comp}
			}
		}
		\caption{Intermediate results for residual coding at a compression rate of roughly 100:1. The block-based inpainting is able to compensate small-scale errors at image edges.}
		\label{fig:res}
	\end{figure*}
	
	\subsection{The Final Codec}
	\label{sec:modeling}
	
	Our final model combines methods explained in the previous sections to obtain a fully inpainting-based video codec. The codec by Andris et al.~\cite{APW16} includes homogeneous diffusion inpainting for intra prediction, the method of Brox et al.~\cite{BBPW04} for inter prediction, quantisation for residual compression, and arithmetic coding for entropy coding. It is able to outperform other inpainting-based codecs, however, it has a limited range of possible compression ratios due to the simple residual compression technique, and real-time decoding has only been shown for a resolution of $854 \times 364$. We aim at extending the compression range, improving quality, and increase decoding speed by consequently filling all submodules of the compression pipeline with suitable methods. 
	
	In a first step, we compute backwards optic flow fields between all frames. As in \cite{APW16}, we employ the method of Brox et al.~\cite{BBPW04}. It produces piecewise smooth motion fields of high quality while still being computationally manageable. Applying their algorithm back to front for all frames in a GOP, we obtain BOFFs for inter prediction that are highly compressible due to their piecewise smoothness but can still yield accurate predictions at motion boundaries. We also tested the much simpler method by Horn and Schunck \cite{HS81}, but got consistently worse results both in terms of reconstruction error and final compression ratio of the video. Thus, it is worthwhile to invest into more advanced methods to acquire accurate flow fields.
	
	The human visual system combines structural with colour information and is less sensitive to errors in the colour domain. Thus, most established codecs transform the input to a colour space with a luma and two chroma channels and compress the chroma channels more strongly. According to this principle, we convert the frame sequence to YUV space with the reversible colour transform (RCT) employed in JPEG2000 and carry out all computations in this domain. In order to realise a stronger compression in the chroma channels, we take half the amount of mask points there compared to the luma channel for all our inpainting methods. More precisely, we use the same mask optimisation technique as for the corresponding luma channel to acquire one inpainting mask for both chroma channels.
	
	Since we only have to reconstruct one intra frame per GOP, we choose the global homogeneous diffusion inpainting method (Section \ref{sec:hominp}) for prediction. This results in a lower prediction error at the cost of an increased runtime. We acquire the corresponding inpainting mask with a rectangular subdivision scheme introduced by Schmaltz et al.~\cite{SPMEWB14}, which allows to concentrate mask points in regions with large reconstruction errors. To further reduce storage, we quantise colour values with a simple uniform quantisation.
	
	For flow fields and residuals, we have more severe restrictions regarding 
	runtime, since we have to recover one of each for every reconstructed frame. 
	Experiments showed that typical BOFFs consist of very large regions with barely 
	changing values. Thus, we use the rectangular subdivision scheme from 
	\cite{SPMEWB14} and assign each region its average value. Residuals tend to 
	contain more structure, since both intra and inter prediction work best on 
	smooth regions. Therefore, we opt for block-based pseudodifferential inpainting 
	with the harmonic operator as described in Section \ref{sec:pseudo}. For every 
	$8 \times 8$ block we again acquire mask positions with rectangular 
	subdivision. The resulting coefficients of the Green's functions in general lie 
	in a larger range than the standard colour values and also attain negative 
	values. Therefore, we map the coefficients to  $[-127,127]$ and apply uniform 
	quantisation centered around zero, also called dead-zone quantisation (see e.g. 
	Chapter 3 in \cite{TM02}). This way, we can ensure that zero coefficients are 
	reconstructed as zero again and thus avoid flickering.
	
	After dead-zone quantisation, we can expect to produce data that have a high probability of being zero or close to zero. Therefore, we adopt an idea from JPEG and define categories which describe larger ranges for exceeding distance of its values from zero. We represent the categories with Huffman codes and store additional bits for the exact value. Finally, we remove remaining redundancies with an entropy coder. The well-known concept of arithmetic coding (AC) by Rissanen \cite{Ri76} and variants thereof are still successfully used in various codecs. However, its problem of requiring costly arithmetic operations has only partly been solved. Collet's Finite State Entropy (FSE) coder \cite{FSE} presents an interesting alternative. It performs on par with arithmetic coding, but is consistenly faster. Our implementation is a lightweight version of FSE specifically designed to fit our video codec structure.
	
	% ------------------------------------------------------------------------------
	% Experiments
	
	\begin{figure}[t]
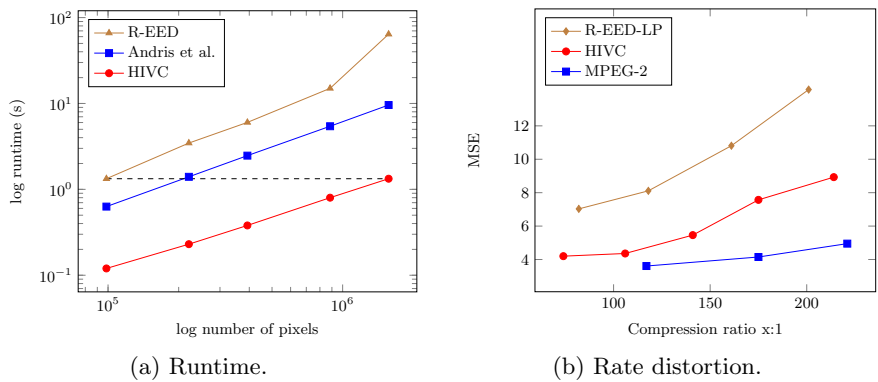

		
		\subfloat[Runtime.]{
			\includestandalone[width=.45\textwidth]{Pics/runtime/runtime}
			\label{fig:runtime}}
		\hfil
		\subfloat[Rate distortion.]{
			\includestandalone[width=.45\textwidth]{Pics/ratedisto/ratedisto}
			\label{fig:ratedisto}}
		
		\caption{Comparison with other codecs in terms of runtime and rate distortion. \textbf{Left:} Runtime depending on the number of pixels in one frame. The dashed line marks real-time decoding with 24 frames per second. The proposed HIVC approach outperforms the other codecs by roughly one order of magnitude. \textbf{Right:} Quality comparisons over several compression ratios. HIVC outperforms R-EED-LP consistently.}
	\end{figure} 
	
	\section{Experiments}
	\label{sec:experiments}
	
	We perform experiments on a sequence of 32 frames of the \emph{Sintel} video by Rosendaal \cite{R11} with an \emph{Intel Xeon CPU W3565@3.20GHz}. 
	
	In order to provide an insight into how our codec's different modules operate, we show intermediate results of the individual coding steps for an intra frame in Figure \ref{fig:intra}. Inter frame processing (Figure \ref{fig:inter}) looks similar, with the difference that we use motion compensation for prediction instead of inpainting.
	
	For decoder run-time comparison, we use the 1080p version of \emph{Sintel}. Figure \ref{fig:runtime} shows results over resolutions ranging from $480\times205$ to $1920\times818$ for our proposed codec HIVC, the codec by Andris et al.\cite{APW16}, and R-EED \cite{PSMMW15}. For the R-EED video codec we have to perform the runtime tests on greyscale videos, but still use colour videos for HIVC and Andris et al., giving R-EED a significant advantage. Our new codec outperforms the other methods by roughly one order of magnitude and is the only codec capable of real-time decoding beyond a resolution of $720\times307$.
	
	In Figure \ref{fig:ratedisto}, we compare the reconstruction quality of our approach with the best possible inpainting-based video compression method that provides an encoder as well as a decoder. Since the R-EED video codec only works on greyvalue videos, we supplement it by the state-of-the-art colour image compression codec R-EED-LP \cite{PKW16}. This is straight-forward since the original codec is completely frame-based, however, the resulting codec is significantly slower. Although it is not yet our goal to beat the established codecs of the MPEG family, we provide results for H.262/MPEG-2 \cite{HPN96} as a point of reference. \emph{Sintel} at a resolution of $960\times409$ serves as test video. We do not include the codec by Andris et al., since it is only able to compress at small compression ratios up to 35:1. In contrast, we are able to reach much larger ratios since our codec consequently integrates inpainting ideas in every prediction and compression step. Especially for BOFF and residual storage, our methods are much more efficient compared to \cite{APW16}. Furthermore, \cite{APW16} can only compete with R-EED-LP on highly textured videos. Here, we pick a sequence with a low amount of texture that is very well suited for R-EED-LP compression, but HIVC is still able to outperform it consistently over several compression ratios. 
	
	% ------------------------------------------------------------------------------
	% Conclusion
	
	\section{Conclusion and Outlook}
	\label{sec:conclusion}
	
	Real-time decoding presents a significant challenge for codecs based on 
	inpainting ideas that so far could not even be reached in Full-HD with 
	parallelisation on GPUs. Our approach is not only the first fully 
	inpainting-based codec to achieve this, but also relies only on a standard CPU. 
	This is possible due to a new pseuodifferential inpainting technique that 
	allows highly efficient reconstruction through fast cosine transforms and 
	carefully selected components for coding and prediction. This allows us to
	outperform previous inpainting-based video codecs \cite{APW16} by one order of 
	magnitude in speed, while simultaneously providing better quality than 
	state-of-the-art inpainting \cite{PKW16}. 
	
	Our significant advances show the high potential of 
	inpainting-based codecs. In future work, we aim at real-time decoding of 4K 
	video by pushing efficient engineering even further and exploiting 
	parallelisation on GPUs. Furthermore, our pseudodifferential inpainting may 
	give an insight into connections between transform-based and inpainting-based 
	methods and inherently incorporates a variety of linear operators. 
	
	\pagebreak
	
	\bibliographystyle{splncs04}
	\bibliography{andris-refs}

\end{document}

%% file: Pics/runtime/runtime.tex
	\begin{tikzpicture}
\begin{loglogaxis}[
legend cell align={left},
legend pos=north west,
xlabel={log number of pixels},
ylabel={log runtime (s)},
]

% R-EED
\addplot[color=brown,mark=triangle*] coordinates {
	(98400,1.33) 
	(221040,3.46)
	(392640,6.02)
	(884160,15)
	(1570560,64.1)
};

% PCS
\addplot[color=blue,mark=square*] coordinates {
	(98400,0.63)
	(221040,1.4)
	(392640,2.47)
	(884160,5.43)
	(1570560,9.59)
};

%	% PCS FSE
%	\addplot[color=black,mark=square] coordinates {
%		(98400,0.49)
%		(221040,1.17)
%		(392640,2.14)
%		(884160,5.1)
%		(1570560,9.16)
%	};

% TCSVT 
\addplot[color=red,mark=*] coordinates {
	(98400,0.12)
	(221040,0.23)
	(392640,0.38)
	(884160,0.8)
	(1570560,1.33)
};

%Real-time
\addplot[dashed,no marks] coordinates {
	(98400,1.33)
	(1570560,1.33)
};

\legend{R-EED,Andris et al.,HIVC}
\end{loglogaxis}
\end{tikzpicture}

%% file: Pics/ratedisto/ratedisto.tex
		\begin{tikzpicture}
	\begin{axis}[
	legend cell align={left},
	legend pos=north west,
	xlabel={Compression ratio x:1},
	ylabel={MSE},
	ymax=19,
	ytick={0,2,4,6,8,10,12}]
	
	% R-EED-LP
	\addplot[color=brown,mark=diamond*] coordinates {
		%(61,5.61) 
		(82,7.03)
		(118,8.11)
		(161,10.81)
		(201,14.17)
		%(289,17.55)
	};
	
	% Proposed no tonal optimisation
	\addplot[color=red,mark=*] coordinates {
		%(42,4.07)
		(74,4.20)
		%(95,4.25)
		(106,4.36)
		(141,5.46)
		(175,7.57)
		%(183,8.06)
		(214,8.93)
	};
	
	% MPEG-2
	\addplot[color=blue,mark=square*] coordinates {
		(117,3.61)
		(175,4.15)
		(221,4.95)
		%(266,5.78)
		%(301,6.45)
	};
	
	\legend{R-EED-LP,HIVC,MPEG-2}
	\end{axis}
	\end{tikzpicture}

%% file: andris-ssvm21.bbl
\begin{thebibliography}{10}
\providecommand{\url}[1]{\texttt{#1}}
\providecommand{\urlprefix}{URL }
\providecommand{\doi}[1]{https://doi.org/#1}

\bibitem{APW16}
Andris, S., Peter, P., Weickert, W.: A proof-of-concept framework for
  {PDE}-based video compression. In: Proc.~2016 Picture Coding Symposium. IEEE
  Computer Society Press, N\"urnberg, Germany (Dec 2016)

\bibitem{AAN88}
Arai, Y., Agui, T., Nakajima, M.: A fast {DCT-SQ} scheme for images. IEICE
  Transactions  \textbf{71}(11),  1095--1097 (Nov 1988)

\bibitem{AWA19}
Augustin, M., Weickert, J., Andris, S.: Pseudodifferential inpainting: The
  missing link between {PDE}- and {RBF}-based interpolation. In: Lellmann, J.,
  Burger, M., Modersitzki, J. (eds.) Scale Space and Variational Methods,
  Lecture Notes in Computer Science, vol. 11603, pp. 67--78. Springer, Cham
  (2019)

\bibitem{BSLRBS11}
Baker, S., Scharstein, D., Lewis, J., Roth, S., Black, M., Szeliski, R.: A
  database and evaluation methodology for optical flow. International Journal
  of Computer Vision  \textbf{92}(1),  1--31 (Oct 2011)

\bibitem{BHR20}
Breu{\ss}, M., Hoeltgen, L., Radow, G.: Towards {PDE}-based video compression
  with optimal masks prolongated by optic flow. Journal of Mathematical Imaging
  and Vision  \textbf{62},  1--13 (2020)

\bibitem{BBPW04}
Brox, T., Bruhn, A., Papenberg, N., Weickert, J.: High accuracy optical flow
  estimation based on a theory for warping. In: Fleet, D., Pajdla, T., Schiele,
  B., Tuytelaars, T. (eds.) Computer Vision-ECCV 2004, vol.~8689, pp. 25--36.
  Springer, Berlin (2004)

\bibitem{Bu14}
Bull, D.: Communicating Pictures: A Course in Image and Video Coding. Academic
  Press, Cambridge, MA (2014)

\bibitem{Ca88}
Carlsson, S.: Sketch based coding of grey level images. Signal Processing
  \textbf{15}(1),  57--83 (Jul 1988)

\bibitem{CM13}
Chen, W., Mied, R.: Optical flow estimation for motion-compensated compression.
  Image and Vision Computing  \textbf{31}(3),  275--289 (Jan 2013)

\bibitem{FSE}
Collet, Y.: Finite state entropy,
  \url{https://github.com/Cyan4973/FiniteStateEntropy}

\bibitem{De94}
Deuflhard, P.: Cascadic conjugate gradient methods for elliptic partial
  differential equations: algorithm and numerical results. In: Keyes, D.E., Xu,
  J. (eds.) Contemporary Mathematics, vol.~180, pp. 29--29. American
  Mathematical Society, Procidence, RI (1994)

\bibitem{DNLKW10}
Doshkov, D., Ndjiki-Nya, P., Lakshman, H., Koppel, M., Wiegand, T.: Towards
  efficient intra prediction based on image inpainting methods. In: Proc.~27th
  Picture Coding Symposium. pp. 470--473. IEEE Computer Society Press, Nagoya,
  Japan (Dec 2010)

\bibitem{GWWB08}
Gali\'c, I., Weickert, J., Welk, M., Bruhn, A., Belyaev, A., Seidel, H.P.:
  Image compression with anisotropic diffusion. Journal of Mathematical Imaging
  and Vision  \textbf{31}(2--3),  255--269 (Jul 2008)

\bibitem{HP01}
Han, S.C., Podilchuk, C.: Video compression with dense motion fields. IEEE
  Transactions on Image Processing  \textbf{10}(11),  1605--1612 (Nov 2001)

\bibitem{HPN96}
Haskell, B.G., Puri, A., Netravali, A.N.: Digital Video: An Introduction to
  MPEG-2. Springer, Berlin (1996)

\bibitem{HPW15}
Hoffmann, S., Plonka, G., Weickert, J.: Discrete {Green}'s functions for
  harmonic and biharmonic inpainting with sparse atoms. In: Tai, X.C., Bae, E.,
  Chan, T.F., Lysaker, M. (eds.) Energy Minimization Methods in Computer Vision
  and Pattern Recognition, Lecture Notes in Computer Science, vol.~8932, pp.
  169--182. Springer, Berlin (2015)

\bibitem{HS81}
Horn, B., Schunck, B.: Determining optical flow. Artificial Intelligence
  \textbf{17},  185--203 (Aug 1981)

\bibitem{JPW20}
Jost, F., Peter, P., Weickert, J.: Compressing flow fields with edge-aware
  homogeneous diffusion inpainting. In: Proc.~45th International Conference on
  Acoustics, Speech, and Signal Processing (ICASSP). pp. 2198--2202. IEEE
  Computer Society Press, Barcelona, Spain (May 2020)

\bibitem{KSFR07}
K\"ostler, H., St\"urmer, M., Freundl, C., R\"ude, U.: {PDE} based video
  compression in real time. Tech. Rep. 07-11, Lehrstuhl f\"ur Informatik 10,
  University Erlangen--N\"urnberg, Germany (2007)

\bibitem{LHX18}
Li, B., Han, J., Xu, Y.: Co-located reference frame interpolation using optical
  flow estimation for video compression. In: Proc.~2018 Data Compression
  Conference. pp. 13--22. IEEE Computer Society Press, Snowbird, {UT} (Mar
  2018)

\bibitem{LSWZ08}
Liu, D., Sun, X., Wu, F., Zhang, Y.Q.: Edge-oriented uniform intra prediction.
  IEEE Transactions on Image Processing  \textbf{17}(10),  1827--1836 (Oct
  2008)

\bibitem{OK13}
Ottaviano, G., Kohli, P.: Compressible motion fields. In: Proc.~2013 IEEE
  Conference on Computer Vision and Pattern Recognition. pp. 2251--2258. IEEE
  Computer Society Press, Oregon, {OH} (Jun 2013)

\bibitem{PKW16}
Peter, P., Kaufhold, L., Weickert, J.: Turning diffusion-based image
  colorization into efficient color compression. IEEE Transactions on Image
  Processing  \textbf{26}(2),  860--869 (2016)

\bibitem{PSMMW15}
Peter, P., Schmaltz, C., Mach, N., Mainberger, M., Weickert, J.: Beyond pure
  quality: {P}rogressive modes, region of interest coding, and real time video
  decoding for {PDE}-based image compression. Journal of Visual Communication
  and Image Representation  \textbf{31}(4),  253--265 (Aug 2015)

\bibitem{Ri76}
Rissanen, J.J.: Generalized {Kraft} inequality and arithmetic coding. IBM
  Journal of Research and Development  \textbf{20}(3),  198--203 (May 1976)

\bibitem{R11}
Roosendaal, T.: Sintel. In: ACM SIGGRAPH 2011 Computer Animation Festival.
  p.~71. New York, NY, USA (2011)

\bibitem{SGPS+95}
Sanchez, V., Garcia, P., Peinado, A.M., Segura, J.C., Rubio, A.J.:
  Diagonalizing properties of the discrete cosine transforms. IEEE Transactions
  on Signal Processing  \textbf{43}(11),  2631--2641 (1995)

\bibitem{SPMEWB14}
Schmaltz, C., Peter, P., Mainberger, M., Ebel, F., Weickert, J., Bruhn, A.:
  Understanding, optimising, and extending data compression with anisotropic
  diffusion. International Journal of Computer Vision  \textbf{108}(3),
  222--240 (Jul 2014)

\bibitem{SW12}
Schmaltz, C., Weickert, J.: Video compression with {3-D} pose tracking,
  {PDE}-based image coding, and electrostatic halftoning. In: Pinz, A., Pock,
  T., Bischof, H., Leberl, F. (eds.) Pattern Recognition, Lecture Notes in
  Computer Science, vol.~7476, pp. 438--447. Springer, Berlin (2012)

\bibitem{SM14}
Strang, G., MacNamara, S.: Functions of difference matrices are {T}oeplitz plus
  {H}ankel. SIAM Review  \textbf{56}(3),  525--546 (Aug 2014)

\bibitem{TBS06}
Tan, T.K., Boon, C.S., Suzuki, Y.: Intra prediction by template matching. In:
  Proc.~2006 IEEE International Conference on Image Processing. pp. 1693--1696.
  IEEE Computer Society Press, Atlanta, {GA}, USA (Oct 2006)

\bibitem{TM02}
Taubman, D.S., Marcellin, M.W. (eds.): {JPEG 2000}: {I}mage Compression
  Fundamentals, Standards and Practice. Kluwer, Boston (2002)

\bibitem{WSK18}
Wu, C.Y., Singhal, N., Kr\"ahenb\"uhl, P.: Video compression through image
  interpolation. In: Ferrari, V., Hebert, M., Sminchisescu, C., Weiss, Y.
  (eds.) Computer Vision-ECCV 2018, vol. 11212, pp. 416--431. Springer, Cham
  (2018)

\bibitem{ZL14}
Zhang, Y., Lin, Y.: Improving {HEVC} intra prediction with {PDE}-based
  inpainting. In: Asia-Pacific Signal and Information Processing Association
  Annual Summit and Conference (APSIPA). IEEE Computer Society Press, Chiang
  Mai, Thailand (Dec 2014)

\end{thebibliography}
